# Diffusion in Evaporating Polymer Solutions: A Model in the Dissipative Formalism of Nonequilibrium Thermodynamics

Siamak. Shams Es-haghi[*]

*Department of Polymer Engineering and Department of Theoretical and Applied Mathematics, The University of Akron, Akron, OH 44325, USA*

**Abstract**

In this paper, diffusion in polymer solutions undergoing evaporation of solvent is modeled as a coupled heat and mass transfer problem with moving boundary condition within the framework of nonequilibrium thermodynamics. The proposed governing equations derived from the fundamental equation of classical thermodynamics using the local equilibrium hypothesis display more complex connection between heat and non-convective mass fluxes than what has been presented in the previous research works. Numerical computations, performed using an explicit finite difference scheme, indicate that the model is able to capture the effect of thermal diffusion in polymer solutions. This effect manifests itself as an increase in local concentration of solvent near warm substrates during solution casting process.

[*] ss172@zips.uakron.edu

# 1. Introduction

Drying of polymer solutions as an interesting example of a nonequilibrium phenomenon is a crucial process which has significant importance in technologies related to painting, coating, manufacturing polymer films and production of electronic devices [1].

In order to have a better control on this process, one should have a good understanding of the physics underlying the problem. Although modeling of this process has been addressed several times [2,3], the proposed governing equations are not theoretically sophisticated enough to describe the whole picture of evolution of the mentioned phenomenon governed by the processes of mass diffusion and heat conduction. These irreversible processes which contribute to entropy production should be studied within the framework of nonequilibrium thermodynamics.

In this paper, it is shown that how local equilibrium hypothesis enables us to use the fundamental equation of classical thermodynamics to derive heat and mass fluxes in the linear region of nonequilibrium thermodynamics taking advantage of Onsager's reciprocity relations.

The proposed governing equations have been solved numerically for a 1-D solution casting problem using an explicit finite difference scheme.

It should be mentioned that we are not concerned about different concentration regimes here; however, it is an interesting topic to be studied in future.

## 2. Model

We consider a non-reacting multi-component mixture in which non-convective mass diffusion and heat conduction occur. The system is considered to be in one phase, far from critical region of phase separation and above its glass transition temperature during the course of evaporation which happens only at surface. We also assume no change in volume upon mixing and consider the system to be in mechanical equilibrium.

The hypothesis of local equilibrium allows the fundamental equation of classical thermodynamics to be valid for every volume element in the system, although the whole system is not in equilibrium [4]. Equation (1) reads the fundamental equation per unit volume of the mixture;

$$de = Tds + \sum_{i=1}^{k} \mu_i dn_i \qquad (1)$$

where $T$, $s$, $e$, $\mu_i$ and $n_i$ are absolute temperature, entropy per unit volume, enthalpy per unit volume, chemical potential of component $i$ and moles of component $i$ per unit volume, respectively.

One can rewrite Eq. (1), using mass concentration of the components and take derivative with respect to time to get Eq. (2);

$$T\frac{ds}{dt} = \frac{de}{dt} - \sum_{i=1}^{k} \frac{\mu_i}{M_i} \frac{dc_i}{dt} \qquad (2)$$

where $c_i$ and $M_i$ are mass concentration and molecular weight of component $i$, respectively.

Time derivatives in Eq. (2) are substantial time derivatives given by Eq. (3).

$$\frac{d}{dt} = \frac{\partial}{\partial t} + v \cdot \nabla \qquad (3)$$

The time derivatives of $e$ and $c_i$ in Eq. (2) are related to the divergence of heat and non-convective mass fluxes and are given by Eqs. (4) and (5), respectively.

$$\frac{de}{dt} = -\nabla \cdot J_q \qquad (4)$$

$$\frac{dc_i}{dt} = -\nabla \cdot J_i \qquad (5)$$

According to Prigogine's theorem, for systems in mechanical equilibrium an arbitrary frame of reference can be chosen [4]. The non-convective mass flux of the component $i$ in this frame of reference which moves with the mean velocity $v$ is given by Eq. (6);

$$J_i = \rho_i \varphi_i (v_i - v) \qquad (6)$$

where $\rho_i$, $\varphi_i$ and $v_i$ are mass density, volume fraction and the velocity of component $i$, respectively and the mean velocity $v$ is expressed as:

$$v = \sum_{i=1}^{k} \varphi_i v_i \qquad (7)$$

It follows from Eqs. (6) and (7) that the fluxes are not independent and Eq. (8) shows their dependency.

$$\sum_{i=1}^{k} \frac{J_i}{\rho_i} = 0 \qquad (8)$$

Rewriting Eq. (2) for a two-component system and replacing $\frac{de}{dt}$ and $\frac{dc_i}{dt}$ with the divergences of the associated fluxes given by Eqs. (4) and (5), Eq. (9) reads:

$$T\frac{ds}{dt} = -\nabla \cdot J_q + \frac{\mu_1}{M_1}\nabla \cdot J_1 + \frac{\mu_2}{M_2}\nabla \cdot J_2 \qquad (9)$$

Hereafter, subscripts 1 and 2 are attributed to solvent and polymer respectively, and $\varphi$ represents the volume fraction of the former.

Equation (9) can be reduced to Eq. (10) after applying the constraint between fluxes given by Eq. (8);

$$T\frac{ds}{dt} = -\nabla \cdot J_q + \mu \nabla \cdot J_1 \tag{10}$$

Where $\mu$ is expressed as:

$$\mu = \frac{\mu_1}{M_1} - \frac{\rho_2}{\rho_1}\frac{\mu_2}{M_2} \tag{11}$$

We can rewrite Eq. (10) by replacing the right hand side with its equivalent and dividing both sides by $T$ which leads to Eq. (12).

$$\frac{ds}{dt} = -\frac{1}{T}\nabla \cdot (J_q - \mu J_1) - \frac{1}{T}J_1 \cdot \nabla \mu \tag{12}$$

The first term of the right hand side of Eq. (12) can be replaced with its equivalent by some manipulation which results in Eq. (13).

$$\frac{ds}{dt} = -\nabla \cdot \left(\frac{J_q - \mu J_1}{T}\right) - \frac{1}{T^2}\nabla T \cdot (J_q - \mu J_1) - \frac{1}{T}J_1 \cdot \nabla \mu \tag{13}$$

We can write Eq. (13) in the following form:

$$\frac{ds}{dt} = -\nabla \cdot J_s + \sigma \tag{14}$$

where $J_s$ given by Eq. (15) is the entropy flux and $\sigma$ expressed by Eq. (16) is the entropy production per unit volume of the system.

$$J_s = \frac{1}{T}(J_q - \mu J_1) \tag{15}$$

$$\sigma = -\frac{1}{T}J_1 \cdot \nabla \mu - \frac{1}{T^2}\nabla T \cdot (J_q - \mu J_1) \tag{16}$$

One can see that $\sigma$ is a bilinear form of fluxes $J_1$ and $(J_q - \mu J_1)$ and forces $-\frac{1}{T}\nabla\mu$ and $-\frac{1}{T^2}\nabla T$.

In the linear region of nonequilibrium thermodynamics, the fluxes can be written in terms of the forces as shown in Eqs. (17) and (18) which in matrix presentation would yield Eq. (19).

$$J_1 = l_{11}\left(-\frac{1}{T}\nabla\mu\right) + l_{12}\left(-\frac{1}{T^2}\nabla T\right) \tag{17}$$

$$J_q - \mu J_1 = l_{21}\left(-\frac{1}{T}\nabla\mu\right) + l_{22}\left(-\frac{1}{T^2}\nabla T\right) \tag{18}$$

$$\begin{pmatrix} J_1 \\ J_q - \mu J_1 \end{pmatrix} = \begin{pmatrix} l_{11} & l_{12} \\ l_{21} & l_{22} \end{pmatrix} \begin{pmatrix} -\frac{1}{T}\nabla\mu \\ -\frac{1}{T^2}\nabla T \end{pmatrix} \tag{19}$$

Entries of matrix $L = (l_{ij})_{2\times 2}$ are Onsager's coefficients and based on Onsager's reciprocity relations, off-diagonal entries of matrix $L$ are identical [5].

In conditions for which linear flux-force relations are valid, entropy production takes the quadratic form given by Eq. (20) [5].

$$\sigma = l_{11}\left(-\frac{1}{T}\nabla\mu\right)^2 + (l_{12} + l_{21})\left(-\frac{1}{T}\nabla\mu\right)\left(-\frac{1}{T^2}\nabla T\right) + l_{22}\left(-\frac{1}{T^2}\nabla T\right)^2 > 0 \tag{20}$$

Matrix $L = (l_{ij})_{2\times 2}$ which satisfies Eq. (20) should be positive definite and to be so, its entries should satisfy the following conditions:

$$l_{11} > 0, \quad l_{22} > 0, \quad l_{11}l_{22} > (l_{12})^2 \tag{21}$$

It is interesting to note that, if we let $\frac{l_{11}}{T} = \alpha$, $\frac{l_{12}}{T^2} = \beta$, $\frac{l_{21}}{T} = \delta$ and $\frac{l_{22}}{T^2} = \gamma$, we can recast the Eqs. (17) and (18) in the same way presented by hydrodynamic calculations of Landau [6].

Doing so, keeping in mind that $\delta$ can be replaced with $\beta T$ because of equality of $l_{12}$ and $l_{21}$, and replacing $\nabla \mu$ with the right hand side of Eq. (22), we will get Eqs. (23) and (24).

$$\nabla \mu = \left(\frac{\partial \mu}{\partial \varphi}\right)_T \nabla \varphi + \left(\frac{\partial \mu}{\partial T}\right)_\varphi \nabla T \tag{22}$$

$$J_1 = -\alpha \left(\frac{\partial \mu}{\partial \varphi}\right)_T \nabla \varphi - \left[\alpha \left(\frac{\partial \mu}{\partial T}\right)_\varphi + \beta\right] \nabla T \tag{23}$$

$$J_q = \left(\mu + \frac{\beta}{\alpha} T\right) J_1 - \left(\gamma - \frac{\beta^2}{\alpha} T\right) \nabla T \tag{24}$$

In order to preserve the positive definiteness of matrix $L$, following conditions should be satisfied:

$$\alpha > 0, \quad \gamma > 0, \quad \beta < \sqrt{\alpha \gamma T^{-1}} \tag{25}$$

After deriving the heat and mass fluxes, the governing equations are simply given by Eqs. (26) and (27);

$$\rho_1 \frac{d\varphi}{dt} = \nabla \cdot \left\{\alpha \left(\frac{\partial \mu}{\partial \varphi}\right)_T \nabla \varphi + \left[\alpha \left(\frac{\partial \mu}{\partial T}\right)_\varphi + \beta\right] \nabla T\right\} \tag{26}$$

$$\rho c_P \frac{dT}{dt} = -\nabla \cdot \left\{\left(\mu + \frac{\beta}{\alpha} T\right) J_1 - \left(\gamma - \frac{\beta^2}{\alpha} T\right) \nabla T\right\} \tag{27}$$

where $\rho$ and $c_P$ are mass density and isobaric specific heat capacity of the solution respectively and are assumed to be constant.

It should be noticed that $\alpha \left(\frac{\partial \mu}{\partial \varphi}\right)_T$, $\left[\alpha \left(\frac{\partial \mu}{\partial T}\right)_\varphi + \beta\right]$ and $\left(\gamma - \frac{\beta^2}{\alpha} T\right)$ are mutual diffusion coefficient $D_M$, thermal diffusion coefficient $D_T$ and thermal conductivity respectively.

In general, phenomenological coefficients $\alpha$, $\beta$ and $\gamma$ can be functions of temperature and concentration. Since we do not have any theoretical knowledge of these functions, the most simplest functions shall be considered for the numerical computations based on the constraints mentioned earlier due to positive definiteness of matrix $L$, dimensional analysis and important asymptotic behavior of mutual and thermal diffusion coefficients at very small concentrations $(\varphi \to 0)$ where the former tends to a finite constant and the latter tends to zero.

To satisfy the constraint $\beta < \sqrt{\alpha \gamma T^{-1}}$, we let $\beta$ be zero and $\alpha$ and $\gamma$ are given by Eqs. (28) and (29). In fact, here we ignore correlation effects between non-convective mass flux $J_1$ and reduced heat flux $J_q - \mu J_1$, and therefore the off-diagonal entries of Onsager matrix, which are given by Green-Kubo relation in terms of integral of time correlation functions of the mentioned fluxes, are reduced to zero.

$$\alpha = \frac{D_0 \rho_1 M_1 \varphi (1-\varphi)}{RT} \tag{28}$$

$$\gamma = \gamma_0 \frac{T_b^{6/5}}{M_1 T_c^{1/6}} \frac{(1-T_r)^{0.38}}{T_r^{1/6}} \tag{29}$$

where $D_0$ and $\gamma_0$ are two constant parameters, $T_b$ and $T_c$ are boiling and critical points of the solvent and $T_r$ is reduced temperature [7].

$\mu$ can be calculated via Eq. (11) while chemical potentials of solvent and polymer are expressed as follows [8]:

$$\mu_1 = \mu_1^0 + RT \left[ \ln(\varphi) + (1-\varphi)\left(1 - \frac{1}{N}\right) + \chi(1-\varphi)^2 \right] \tag{30}$$

$$\mu_2 = \mu_2^0 + RT \left[ \ln(1-\varphi) + \varphi(1-N) + \chi N \varphi^2 \right] \tag{31}$$

$\mu_1^0$, $\mu_2^0$, $R$, $\chi$ are chemical potential of pure solvent, chemical potential of the pure liquid polymer, universal gas constant, Flory-Huggins interaction parameter, and $N$ is the ratio of the molar volumes of polymer and solvent.

Assuming molecular weight of polymer to be very high as compared to that of solvent, $\mu$ can be expressed by Eq. (32).

$$\mu = \frac{\mu_1^0}{M_1} + \frac{RT}{M_1}\left[\ln(\varphi) + 1 + \chi(1-2\varphi)\right] \tag{32}$$

The chemical potential of pure solvent is assumed to be a linear function of temperature and is given by Eq. (33);

$$\mu_1^0(T) = a + b(T - T_0) \tag{33}$$

where $a$ is the chemical potential at temperature $T_0$ and $b$ is the temperature coefficient.

## 3. Initial and boundary conditions

In this section the proposed governing equations will be used for numerical computations of drying process of a thin polymer solution film on a glass substrate while the whole system is exposed to a laminar flow of air with temperature $T_\infty$ (Fig. 1). The film is thin in the sense that its dimensions in the plane of surface, which is perpendicular to $y$-axis, are large compared to the thickness, and hence the diffusion is a 1-D process to a good approximation.

For each governing equation, two boundary conditions and an initial condition are needed to solve the problem. The initial conditions are fulfilled by starting with uniform concentration and temperature profiles expressed as:

$$\varphi(y, t=0) = \varphi_0 \tag{34}$$

$$T(y, t=0) = T_0 \tag{35}$$

For numerical computations $\varphi_0$ and $T_0$ are considered to be 0.8 and $298.15K$, respectively.

Boundary conditions at surface of the solution are given by Eqs. (36) and (37);

$$(J_1)_{lab, w(t)} - c_1(w(t), t) \frac{dw(t)}{dt} = C_s \tag{36}$$

$$(J_q)_{lab, w(t)} = \langle h \rangle (T - T_\infty) + \Delta H_v C_s \tag{37}$$

where $w(t)$, $C_s$, $\langle h \rangle$ and $\Delta H_v$ are the thickness of the film, evaporating solvent flux, averaged heat transfer coefficient and heat of evaporation of solvent, respectively. Subscript 'lab' denotes laboratory frame of reference.

Evaporating solvent flux can be calculated via Eq. (38);

$$C_s = \langle K_c \rangle (c_{sw} - c_{s\infty}) \tag{38}$$

where $\langle K_c \rangle$ and $c_{sw}$ are averaged mass transfer coefficient and mass concentration of solvent at liquid-air interface. $c_{s\infty}$ is mass concentration of solvent far from the interface, which will be assumed to be negligible.

Averaged transfer coefficients for a laminar flow of uniform velocity $v_\infty$ over a flat plate with the length of $x$ in the direction of flow are as follows [9]:

$$\langle h \rangle = \frac{\langle Nu \rangle k_{air}}{x} \tag{39}$$

$$\langle K_c \rangle = \frac{\langle Sh \rangle D_{AB}}{x} \tag{40}$$

where $k_{air}$, $D_{AB}$, $\langle Nu \rangle$ and $\langle Sh \rangle$ are thermal conductivity of air, diffusion coefficient of solvent into air, the averaged Nusselt and Sherwood numbers, respectively. These numbers can be approximated by Eqs. (41) and (42) [9];

$$\langle Nu \rangle = 0.664 \, \mathrm{Re}^{1/2} \, \mathrm{Pr}^{1/3} \tag{41}$$

$$\langle Sh \rangle = 0.664 \, \mathrm{Re}^{1/2} \, Sc^{1/3} \tag{42}$$

where Re, Pr and $Sc$ are Reynolds, Prandtl and Schmidt numbers.

Assuming ideal gas behavior for solvent at liquid-air interface, $c_{sw}$ can be calculated through the equation of state of ideal gas.

$$c_{sw} = \frac{P_S M_1}{R T_\infty} \tag{43}$$

According to Flory – Huggins theory, the ratio of vapor pressure of solvent in polymer solution $P_S$ to the vapor pressure of pure solvent $P_S^*$ is given by [8]:

$$\frac{P_S}{P_S^*} = \exp\left[\ln(\varphi) + (1-\varphi) + \chi(1-\varphi)^2\right] \tag{44}$$

The vapor pressure of pure solvent at a given temperature $T$ can be estimated from a known value $P_{S0}^*$ at a temperature $T_0$ by using the Clausius – Clapeyron equation (Eq. (45));

$$\ln\left(\frac{P_S^*}{P_{S0}^*}\right) = -\frac{\Delta H}{R}\left(\frac{1}{T} - \frac{1}{T_0}\right) \tag{45}$$

where $\Delta H$ is molar heat of evaporation.

It should be noted that mass and heat fluxes have been derived in a frame of reference which moves with the velocity given by Eq. (7); hence boundary conditions should be expressed in the same frame of reference. After deriving the equation for moving boundary $w(t)$, it will be shown that the fluxes at the boundaries are the same in both laboratory and moving frame of references. Boundary conditions at substrate for mass and heat equations are as follows:

$$(J_1)_{lab} = 0 \tag{46}$$

$$(J_q)_{lab} = -k\frac{\partial T_{glass}}{\partial y} \tag{47}$$

Where $k$ and $T_{glass}$ are thermal conductivity and absolute temperature of glass.

Transient heat conduction in glass substrate along with associated initial and boundary conditions are given by Eqs. (48-50);

$$\frac{\partial T_{glass}}{\partial t} = \kappa \frac{\partial^2 T_{glass}}{\partial y^2} \tag{48}$$

$$T_{glass}(y, t=0) = T_0 \tag{49}$$

$$\langle h \rangle (T_\infty - T_{glass}) = -k\frac{\partial T_{glass}}{\partial y} \tag{50}$$

where $\kappa$ is the thermal diffusivity of glass.

In this problem the location of surface of the solution $w(t)$ is not known a priori and it must be determined as part of the problem. The moving boundary equation can be calculated with regard to conservation of polymer mass during the drying process which would yield Eq. (51).

$$\frac{d}{dt}\int_0^{w(t)} \varphi_2(y,t)dy = 0 \tag{51}$$

Equation (51) can be expanded in terms of volume fraction of solvent as follows:

$$(1-\varphi_w)\frac{dw(t)}{dt} - \int_0^{w(t)} \frac{\partial \varphi}{\partial t}dy = 0 \tag{52}$$

where $\varphi_w$, represents volume fraction of solvent at the moving boundary.

Replacing $\frac{\partial \varphi}{\partial t}$ with $-\frac{1}{\rho_1}\frac{\partial (J_1)_{lab,w(t)}}{\partial y}$ would lead us to:

$$(1-\varphi_w)\frac{dw(t)}{dt} = -\frac{1}{\rho_1}(J_1)_{lab,w(t)} \tag{53}$$

The equation of moving boundary can be found using Eqs. (36) and (53).

$$\frac{dw(t)}{dt} = -\frac{C_s}{\rho_1} \tag{54}$$

The non-convective mass flux of the component $i$ in the laboratory frame of reference is given by Eq. (55).

$$(J_i)_{lab} = c_i v_i \tag{55}$$

The non-convective mass flux of component $i$ in moving and laboratory frame of references can be related by:

$$J_i = (J_i)_{lab} - c_i v \tag{56}$$

The velocity of moving frame of reference at surface of the solution $v$ can be shown by Eq. (57).

$$v = v_1 \varphi + \frac{dw(t)}{dt}(1-\varphi) \tag{57}$$

The velocity of polymer at surface, $\frac{dw(t)}{dt}$, can be calculated by writing Eq. (36) for mass flux of polymer. Writing Eq. (56) for solvent and using Eqs. (54) and (57), one can show that mass flux at surface of the film is the same in both laboratory and moving frame of references. The same condition would be found for the mass flux at substrate using Eqs. (46), (55) and (56).

In case of heat flux, it can be shown that for slow processes such as diffusion, the heat flux is invariant to a change in reference velocity [10] which means that Eq. (58) holds for both boundaries.

$$(J_q)_{lab} = J_q \tag{58}$$

## 4. Computation results and discussion

We solve Eqs. (26) and (27) for a 1-D problem under aforementioned initial and boundary conditions using an explicit finite difference scheme.

The aim of numerical computations is to visualize the trend of variation of concentration and temperature which will be predicted by the proposed governing equations based on the assumptions we have made. The physical properties used for these computations are given in the appendix.

Figures 2-4 show concentration and temperature profiles at the onset of evaporation for different values of $T_\infty$. By concentration we mean volume fraction of the solvent $\varphi$ and by temperature non-dimensionalized temperature $\theta$ given by:

$$\theta = \frac{T}{T_0} \qquad (59)$$

The important point which is barely discernable in the profiles is the variation of volume fraction in the vicinity of the substrate. Upon applying boundary conditions, which would perturb the initial uniform profiles, concentration and temperature gradients will be developed in the system. Based on experimental data, polymers often have positive thermal diffusion coefficient and therefore tend to migrate to cold regions when are exposed to a temperature gradient [11]. Keeping this in mind, an increase in the local concentration of solvent near a warm substrate would be expected. To examine the result of numerical computations, the calculated values of $\varphi$ and $\theta$ for the first three seconds of evaporation in different ambient temperatures are tabulated in Tables 1-3.

Obviously, in Tables 1 and 2 which are related to the situations in which substrate is exposed to higher temperatures than the initial temperature, an increase in the volume fraction of solvent is observed and values of $\varphi$ in the grid points near the substrate increase by passing time which is consistent with the corresponding increase in $\theta$ values. Reduction in $\varphi$ observed in second 3 in Table 2, is due to the departure of solvent from evaporating boundary. This gradual reduction is more pronounced at higher ambient temperatures, as could be seen by making a comparison among Figs. 2-4, and evidently its effect would be sensed faster near the substrate.

Numerical results given by Table 3 confirm that when ambient temperature is identical to the initial temperature, $\varphi$ would not increase near the substrate. Therefore, the calculated values are in accord with the expected effect of thermal diffusion. As already mentioned, we assumed that $\beta$ is zero and therefore in computations we only consider the contribution of temperature dependency of $\mu$ to $D_T$ which is given by Eq. (60).

$$D_T = \alpha \left( \frac{\partial \mu}{\partial T} \right)_\varphi \tag{60}$$

Variations of surface temperature and thickness as functions of time during the drying process when $T_\infty = 318.15 K$ and initial thickness is $200 \mu m$ are shown in Fig. 5.

The concentration and temperature profiles associated with Fig. 5, are given by Figs. 6 and 7, respectively.

## 5. Conclusions

Mass diffusion and heat conduction in polymer solutions under evaporation were studied within the framework of nonequilibrium thermodynamics. The derived governing equations indicate that non-convective mass flux contributes to the heat equation which cannot be seen in previous research studies. Mutual and thermal diffusion coefficients and thermal conductivity of polymer solution are presented in terms of phenomenological coefficients for which we have some theoretical constraints. To perform numerical computations, these coefficients were approximated in a way that the theoretical constraints and the asymptotic behaviors of mutual and thermal diffusion coefficients are satisfied. Despite the assumptions we have made, numerical results indicate that the model can capture the effect of thermal diffusion in the system and the trends of variations of film thickness and surface temperature of the film are consistent with what is expected to be observed.

**Acknowledgment**

The author is grateful to Professor A. I. Leonov for fruitful discussions.

**Appendix**

$\rho_1 = 779 \, kg.m^{-3}$

$M_1 = 84.16 \, g.mol^{-1}$

$\rho_2 = 1050 \, kg.m^{-3}$

$\chi = 0.35$

$R = 8.314 \, J.K^{-1}.mol^{-1}$

$T_b = 353.9 \, K$

$T_c = 553.4 \, K$

$a = 26.83 \times 10^3 \, J.mol^{-1}$

$b = -204.10 \, J.K^{-1}.mol^{-1}$

$D_0 = 10^{-8} \, m^2.\sec^{-1}$

$\gamma_0 = 0.031 \, W.m^{-1}.K^{-1}$

$c_P = 1.251 \times 10^3 \, J.kg^{-1}.K^{-1}$

$\Delta H_v = 0.38 \times 10^6 \, J.kg^{-1}$

$\Delta H = 32 \times 10^3 \, J.mol^{-1}$

$v_\infty = 1.2 \, m.\sec^{-1}$

**Captions**

**Figure captions:**

Fig. 1. Schematic representation of the solvent evaporation process

Fig. 2. Concentration and temperature profiles at the onset of evaporation when $T_\infty = 298.15K$ ( $x$ represents non-dimensionalized thickness)

Fig. 3. Concentration and temperature profiles at the onset of evaporation when $T_\infty = 318.15K$ ( $x$ represents non-dimensionalized thickness)

Fig. 4. Concentration and temperature profiles at the onset of evaporation when $T_\infty = 338.15K$ ( $x$ represents non-dimensionalized thickness)

Fig. 5. Surface temperature and thickness of the film as functions of time ( $T_\infty = 318.15K$ )

Fig. 6. Concentration profiles during the course of drying of polymer solution ( $T_\infty = 318.15K$ )

Fig. 7. Temperature profiles during the course of drying of polymer solution ( $T_\infty = 318.15K$ )

**Table captions:**

Table 1. Calculated values of solvent volume fraction $\varphi$ and non-dimensionalized temperature $\theta$ at five grid points for $T_\infty = 318.15K$ (Grid point 1 represents substrate)

Table 2. Calculated values of solvent volume fraction $\varphi$ and non-dimensionalized temperature $\theta$ at five grid points for $T_\infty = 338.15K$ (Grid point 1 represents substrate)

Table 3. Calculated values of solvent volume fraction $\varphi$ and non-dimensionalized temperature $\theta$ at five grid points for $T_\infty = 298.15K$ (Grid point 1 represents substrate)

**Figures:**

Fig. 1.

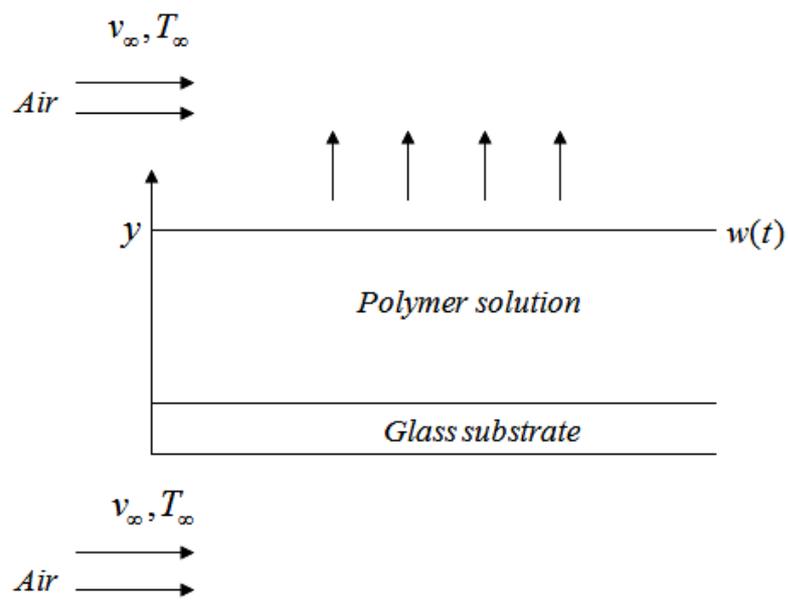

Fig. 2.

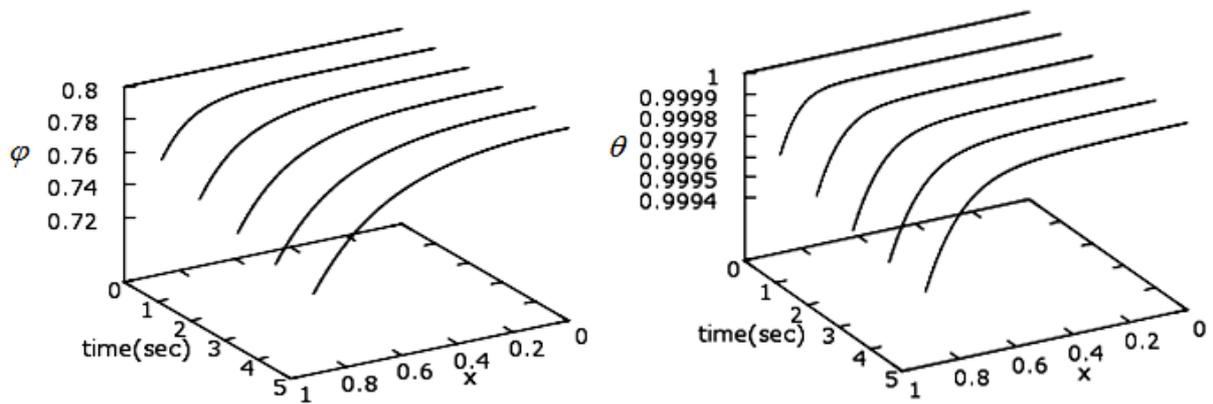

Fig. 3.

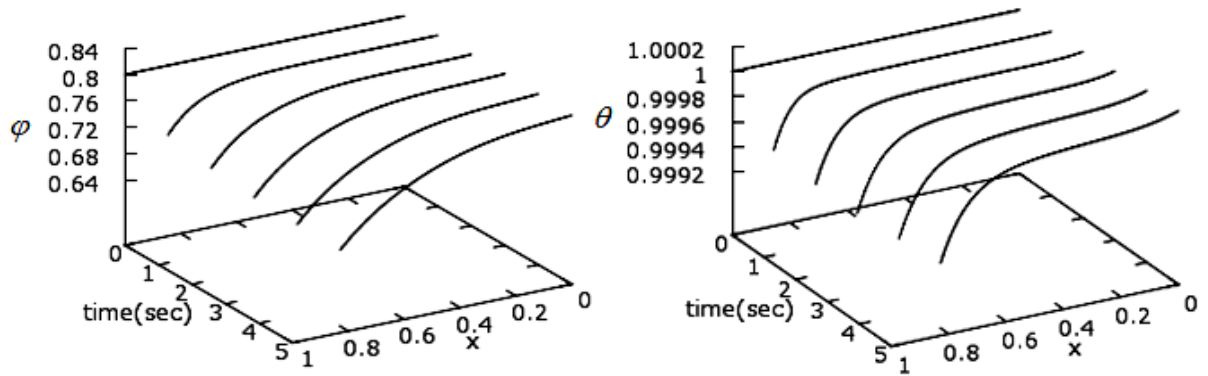

Fig. 4.

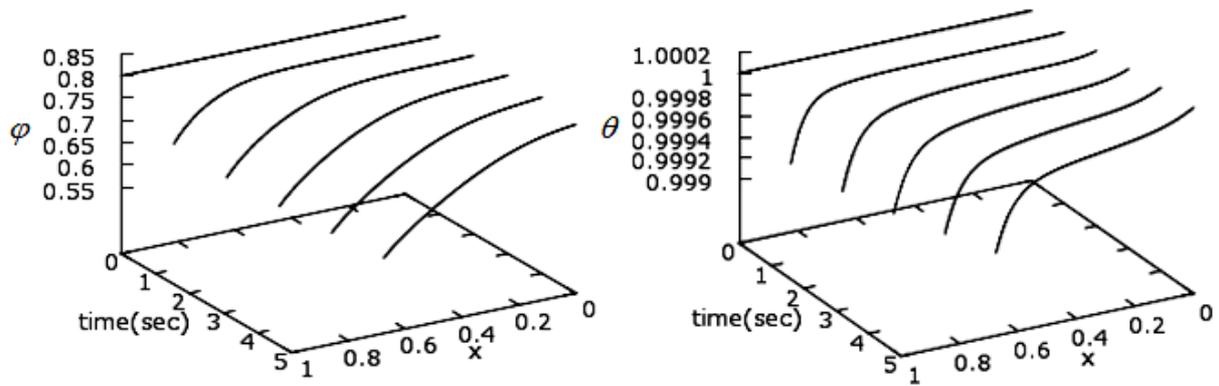

Fig. 5.

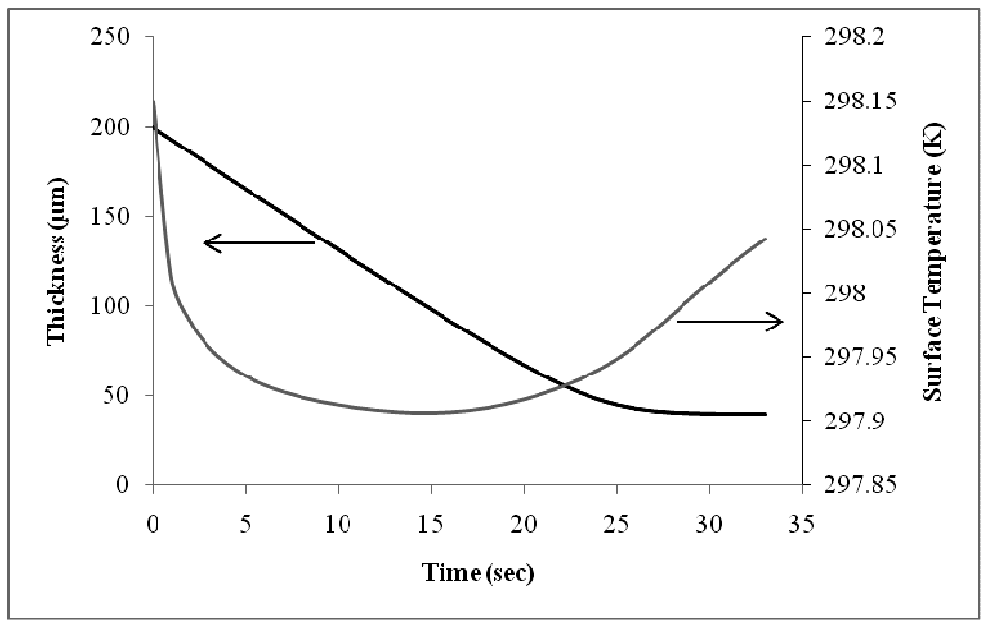

Fig. 6.

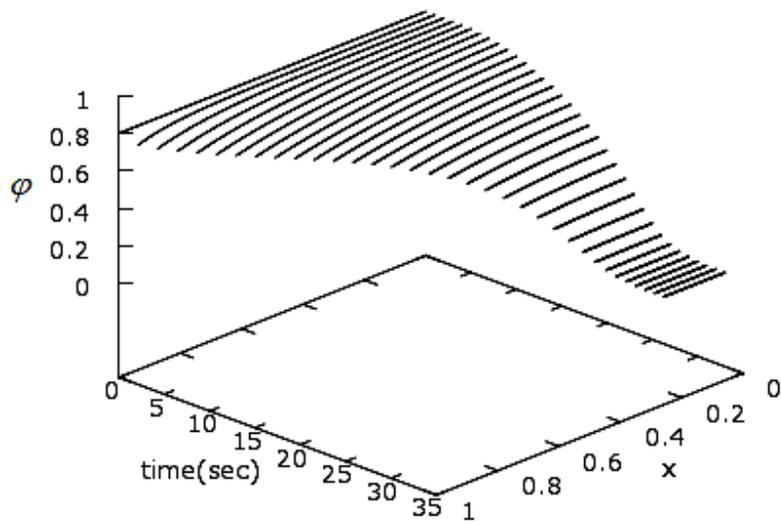

Fig. 7.

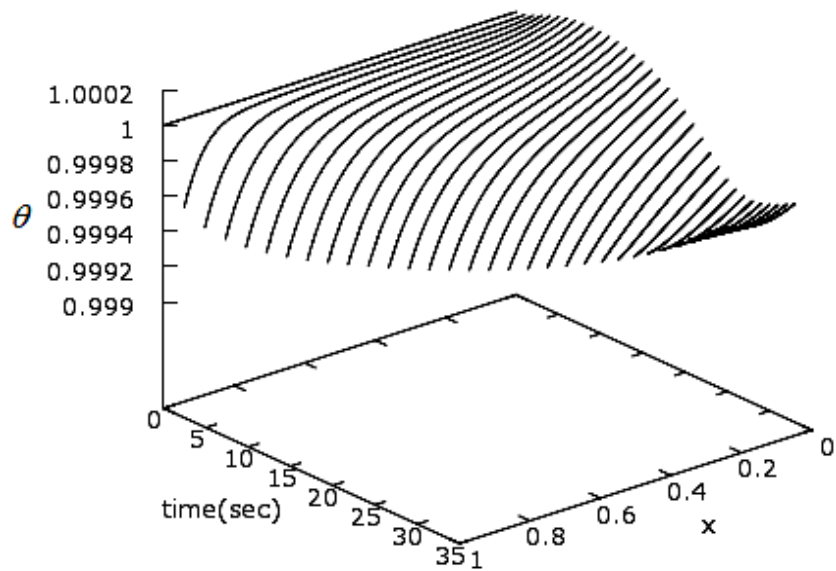

**Tables:**

Table 1.

| Time = 1 sec | | | Time = 2 sec | | | Time = 3 sec | | |
|---|---|---|---|---|---|---|---|---|
| Grid point | $\varphi$ | $\theta$ | Grid point | $\varphi$ | $\theta$ | Grid point | $\varphi$ | $\theta$ |
| 1 | 0.800149035 | 1.00000477 | 1 | 0.800680339 | 1.00002282 | 1 | 0.800955006 | 1.00004438 |
| 2 | 0.800098586 | 1.00000359 | 2 | 0.800533054 | 1.00001943 | 2 | 0.800739982 | 1.00003947 |
| 3 | 0.80006074 | 1.00000268 | 3 | 0.800406284 | 1.00001647 | 3 | 0.800543352 | 1.000035 |
| 4 | 0.80003294 | 1.00000198 | 4 | 0.800297846 | 1.00001389 | 4 | 0.800363623 | 1.00003093 |
| 5 | 0.800013032 | 1.00000144 | 5 | 0.80020567 | 1.00001165 | 5 | 0.800199314 | 1.00002724 |

Table 2.

| Time = 1 sec | | | Time = 2 sec | | | Time = 3 sec | | |
|---|---|---|---|---|---|---|---|---|
| Grid point | $\varphi$ | $\theta$ | Grid point | $\varphi$ | $\theta$ | Grid point | $\varphi$ | $\theta$ |
| 1 | 0.800297689 | 1.00000952 | 1 | 0.80133574 | 1.0000456 | 1 | 0.801279853 | 1.00008802 |
| 2 | 0.800199727 | 1.00000725 | 2 | 0.801059115 | 1.00003924 | 2 | 0.800889376 | 1.00007916 |
| 3 | 0.800125512 | 1.00000547 | 3 | 0.800818177 | 1.00003364 | 3 | 0.800524402 | 1.000071 |
| 4 | 0.800070374 | 1.00000408 | 4 | 0.800609312 | 1.00002871 | 4 | 0.800182695 | 1.0000635 |
| 5 | 0.800030354 | 1.00000301 | 5 | 0.800429083 | 1.0000244 | 5 | 0.799862018 | 1.00005662 |

Table 3.

| Time = 1 sec | | | Time = 2 sec | | | Time = 3 sec | | |
|---|---|---|---|---|---|---|---|---|
| Grid point | $\varphi$ | $\theta$ | Grid point | $\varphi$ | $\theta$ | Grid point | $\varphi$ | $\theta$ |
| 1 | 0.79999999 | 1 | 1 | 0.799989296 | 0.999999988 | 1 | 0.799853599 | 0.999999834 |
| 2 | 0.79999999 | 1 | 2 | 0.79998918 | 0.999999987 | 2 | 0.79985277 | 0.999999828 |
| 3 | 0.799999989 | 1 | 3 | 0.799988888 | 0.999999986 | 3 | 0.799850806 | 0.999999821 |
| 4 | 0.799999987 | 1 | 4 | 0.799988414 | 0.999999986 | 4 | 0.799847688 | 0.999999813 |
| 5 | 0.799999985 | 1 | 5 | 0.79998775 | 0.999999984 | 5 | 0.799843393 | 0.999999805 |